\documentclass[runningheads]{llncs}
\usepackage{graphicx}
\usepackage{subcaption}
\usepackage{xcolor}
\usepackage{amssymb}
\usepackage{multirow}
\usepackage{graphicx}
\usepackage{makecell}

\begin{document}
\title{audioLIME: Listenable Explanations Using Source Separation
}
%
%
\author{Verena Haunschmid\inst{1}\and
Ethan Manilow\inst{3}\and
Gerhard Widmer\inst{1,2}}
\authorrunning{V. Haunschmid et al.}
%
\institute{Institute of Computational Perception, Johannes Kepler University, Linz, Austria \email{verena.haunschmid@jku.at}\orcidID{0000-0001-5466-7829}\and
LIT Artificial Intelligence Lab, Johannes Kepler University, Linz, Austria
\and
Interactive Audio Lab, Northwestern University, Evanston, IL, USA
}
\maketitle              
\begin{abstract}
\vspace{-0.5cm}
Deep neural networks (DNNs) are successfully applied in a wide variety of music information retrieval (MIR) tasks but their predictions are usually not interpretable.
We propose \textit{audioLIME}, a method based on Local Interpretable Model-agnostic Explanations (LIME), extended by a musical definition of locality.
The perturbations used in LIME are created by switching on/off components extracted by source separation which makes our explanations listenable.
We validate audioLIME on two different music tagging systems and show that it produces sensible explanations in situations where a competing method cannot. 

\end{abstract}

\section{Introduction}
Deep neural networks (DNNs) are used in a wide variety of music information retrieval (MIR) tasks. While they generally achieve great results according to standard metrics, it is hard to interpret how or why they determine their output. This can lead to situations where a network does not learn what its designers intend. One goal of the field of interpretable machine learning is to provide tools for practitioners that push towards making the decisions of opaque models understandable. The field of MIR has many stakeholders--from individual musicians to entire corporations--all of which must be able to trust DNN systems.

A promising approach to this problem is Local Interpretable Model-agnostic Explanations (LIME)~\cite{Ribeiro0G16}, which produces explanations of predictions from an arbitrary model \textit{post-hoc} by perturbing interpretable components around an input example and fitting a small, surrogate model to explain the original model's prediction. Previous attempts at adopting LIME for MIR tasks have used rectangular regions of a spectrogram for explanations~\cite{MishraSD17SoundLIME}.
This ignores two defining characteristics of audio data: 1) the lack of occlusion of overlapping sounds and, 2) all parts of a single sound might \textit{not} be contiguous on a spectrogram.

In this work, we introduce \textit{audioLIME}, an extension of LIME that
preserves fundamental aspects of audio so 
explanations are \textit{listenable}. To achieve this we propose a new notion of ``locality'' based on estimates from source separation algorithms. We evaluate our method on music tagging systems by feeding the explanation back into the tagger and 
seeing if the prediction changes.
Using this technique, we show that our method is able to explain predictions from a waveform-based music tagger, which previous methods cannot do.
We also provide illustrative examples of listenable explanations from our system.

\section{audioLIME}

\begin{figure}[t]
    \centering
    \includegraphics[width=\linewidth]{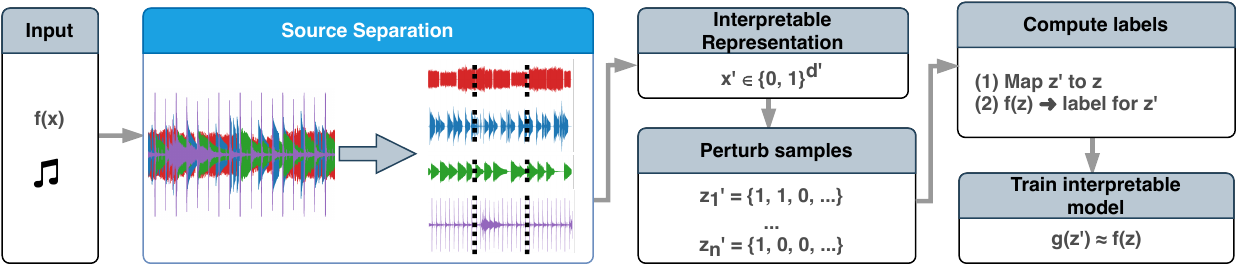}
    \caption{audioLIME closely follows the general LIME pipeline. The key is the use of source estimates (blue box). A source separation algorithm decomposes input audio into $d' = C \times \tau$ interpretable components ($C$ sources, $\tau$ time segments).}
    \label{fig:audioLIME}
\end{figure}

audioLIME is based on the LIME~\cite{Ribeiro0G16} framework and extends its definition of locality for musical data by defining a new way of deriving an interpretable representation. For an input value $x$ to an arbitrary, black-box model $f$, LIME first defines a set of $d'$ interpretable features that can be turned on and off, $x' \in \{0, 1\}^{d'}$. These features are perturbed and represented as a set of binary vectors $z_n'$ that an interpretable, surrogate model trains on. This surrogate model matches the performance of the black-box model around $x$ and is able to reveal which of its interpretable features the black-box model relies on. 

The key insight of audioLIME is that \textit{interpretability} with respect to audio data should really mean \textit{listenability}. Whereas previous approaches applied techniques from the task of image segmentation to spectrograms, we propose using \textit{source separation estimates} as interpretable representations. This gives audioLIME the ability to train on interpretable \textit{and} listenable features.~\footnote{Python package available at: \url{https://github.com/CPJKU/audioLIME}
}




The single-channel source separation problem is formulated as estimating a set of $C$ sources, $S_1, ..., S_c$, when only given access to the mixture $M$ from which the sources are constituents. We note that this definition, as well as audioLIME, is agnostic to the input representation (e.g., waveform, spectrogram, etc) of the audio.
We use these $C$ estimated sources of an input audio as our interpretable components (e.g. \{\textit{piano, drums, vocals, bass}\}). Mapping $z' \in \{0, 1\}^C$ to $z$ (the input audio) is performed by mixing all present sources. For example $z' = \{0, 1, 0, 1\}$ results in a mixture only containing estimates of drums and bass.
The relation of this approach to the notion of \textit{locality} as used in LIME lies in the fact that samples perturbed in this way will in general still be perceptually similar (i.e., recognized by a human as referring to the same audio piece). This system is shown in Figure~\ref{fig:audioLIME}. In addition to source separation, we also segment the audio into $\tau$ temporal segments, resulting in $C \times \tau$ interpretable components. 





%

\section{Experiments}

We analyze two music tagging models~\cite{won2020eval}: FCN~\cite{Choi16FCN}, which inputs a 29 second spectrogram, and SampleCNN~\cite{Lee2017SampleCNN}, which inputs a 3.69 second waveform. Both models were trained on the MillionSongDatatset (MSD)~\cite{Bertin-MahieuxEWL11MSD}.
The LIME explanation model used is a linear regression model trained with l2 regularization on $2^{14}$ samples.
We use Spleeter~\cite{spleeter2019} as the source separation system.

\begin{figure}[t]
    \centering
    \includegraphics[width=0.95\linewidth]{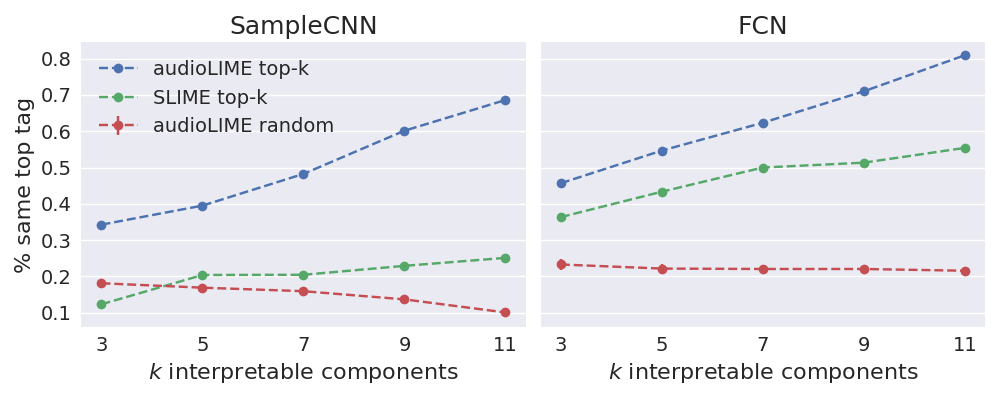}
    \caption{
    Percentage of explanations that produced the same tag as the original input using $k$ interpretable components for two music tagging systems. audioLIME (blue) produces better explanations than SLIME \cite{MishraSD17SoundLIME} (green) and the baseline (red).
    }
    \label{fig:results}
\end{figure}

\vspace{-0.15cm}
\paragraph{Quantitative Results} To verify that the explanations truly explain the model's behaviour we perform a simple experiment. If the explanation explains the model's behaviour we expect the tagger to be able to make the same prediction when only passing the top $k$ selected components, and a different prediction otherwise.\footnote{Experiment code:
\url{https://github.com/expectopatronum/mml2020-experiments/}
}

We randomly picked 100 examples from the MSD test set, 20 for each of the 5 most common tags (rock, pop, alternative, indie, electronic). For each example we create several explanations (3/song for FCN, 16/song for SampleCNN) for the top predicted tag. 
We compare two explanation systems, using the the top $k$ components in each explanation from either audioLIME or SLIME~\cite{MishraSD17SoundLIME}. 
As a baseline, we compare the prediction each tagger makes on $k$ randomly selected components where audioLIME surrogate models have a positive linear weight.

Figure~\ref{fig:results} shows that even when using only a fraction of the components, the tagger makes the same prediction more often with audioLIME than with SLIME or the baseline.
Importantly, because audioLIME's explanations emphasize listenability, they are invariant to the input audio representation of the model, and thus it is able to provide better explanations than SLIME, which does not have the same flexibility.
This indicates there is a whole class of waveform-based models that SLIME is unsuited for, but audioLIME still works well.

\vspace{-1.0cm}
\paragraph{Qualitative Results} Because the explanations audioLIME makes are source estimates, it is possible to listen to and make sense of them. To illustrate this, we selected two examples of explanations of a prediction made by FCN.\footnote{\url{https://soundcloud.com/veroamilbe/sets/mml2020-explanation-example}} In the first example, FCN predicted the tag ``female vocalist'' and, indeed, the top 3 selected audioLIME components are the separated vocals with a female singer. In the second case, FCN predicted the tag ``rock'', and in the top audioLIME components we can hear a driving drumset and a distorted guitar, both of which are associated with rock music. In these cases, we can be confident that our music tagging network has learned the correct concepts for these tags, and thus increases our trust in the black-box FCN model.


\section{Conclusion}

In this work we presented audioLIME, a system that uses source separation to produce \textit{listenable} explanations. We demonstrated an experiment that showed how audioLIME can produce explanations that create trustworthy predictions from music tagging systems that use waveforms or spectrograms as input. We also showed two illustrative examples of explanations from audioLIME. One of the shortcomings of audioLIME is its dependency on a source separation system, which only works with a limited number of source types and may introduce artifacts. However, we note that audioLIME is agnostic to the source separation system, and thus audioLIME is compatible with future work in that space.

\section{Acknowledgements}

This research is supported by the European Research Council (ERC) under the European Union’s Horizon 2020 research and innovation programme, grant agreement No 670035 (``Con Espressione'').

\bibliographystyle{splncs04}
\bibliography{paper}
\end{document}